\documentclass[pre,twocolumn,groupedaddress,floatfix]{revtex4-1}
\usepackage{amssymb,amsmath}
\usepackage{graphicx}
\usepackage{subfigure}
\usepackage[english]{babel} 
\usepackage{float}
\usepackage{color}

\usepackage{lipsum}
\usepackage [latin1]{inputenc}
\begin{document}
\newcommand{\be}{\begin{equation}}
\newcommand{\ee}{\end{equation}}
\newcommand{\rojo}[1]{\textcolor{red}{#1}}

\title{A fractional Anderson model }

\author{Mario I. Molina}
\affiliation{Departamento de F\'{\i}sica, Facultad de Ciencias, Universidad de Chile, Casilla 653, Santiago, Chile}

\date{\today }

\begin{abstract} 
We examine the interplay between disorder and fractionality in a one-dimensional tight-binding Anderson model. In the absence of disorder, we observe that the two lowest energy eigenvalues detach themselves from the bottom of the band, as fractionality $s$ is decreased, becoming completely degenerate at $s=0$, with a common energy equal to a half bandwidth, $V$. The remaining $N-2$ states become completely degenerate forming a flat band with energy equal to a bandwidth, $2V$. Thus, a gap is formed between the ground state and the band. 
In the presence of disorder and for a fixed disorder width, a decrease in $s$ reduces the width of the point spectrum while for a fixed $s$, an increase in disorder increases the width of the spectrum. For all disorder widths, the average participation ratio decreases with $s$ showing a tendency towards localization. However, the average mean square displacement (MSD) shows a hump at low $s$ values, signaling the presence of a population of extended states, in agreement with what is found in long-range hopping models.

\end{abstract}

\maketitle
\noindent
{\bf 1.\ Introduction}.\\
The propagation of excitations in a medium is one of the most studied problems in physics and engineering, either in the classical as well as in the quantum domain. Admittedly, an understanding of the mechanisms that determine heat, mass, charge, momentum, or energy  transport allows, in principle, the management and steering of these signals/excitations between two locations inside a medium, which is of obvious technological importance. The properties of the medium play here  an important role. For simple systems, described by some type of wave equation, it has been recognized that transport is inhibited in the presence of disorder. For instance, in the case of a tight-binding model for an electron propagation inside a crystal, it was shown by Anderson\cite{anderson} that in 1D electron transport is completely inhibited in the presence of any amount of disorder (although some exceptions have been found for the case with long-range coupling. See ref\cite{adame}). The same is true for 2D, while for 3D  a mobility edge is present\cite{mobility}. These phenomena are collectively known as Anderson Localization (AL). It was recognized that the underlying mechanism for AL is the coherent wave interference from the disordered medium, which implies that AL is a rather general phenomenon for all all systems that can display wave-like behavior, such as atomic physics\cite{atomic2, atomic3}, optics\cite{optics1, optics2, optics3,optics4}, plasmonics devices\cite{plasma_anderson}, acoustics\cite{sound,sound2}, spin systems\cite{spin,spin2}, Bose-Einstein condensates\cite{BE1, BE2}, elastic waves\cite{elastic}, among others. The evolution equations that govern the dynamics of excitations in these systems are usually  based on local time and space derivatives of integer order.

On the other hand, the field of fractional calculus has increased its visibility in recent years. It consists of an extension of standard calculus to include derivatives of fractional order. Its beginning dates back to a correspondence between Leibniz and L'Hopital who wondered about a possible extension of the normal integer derivative to non-integer derivates. This would give meaning to the question: `What is the half derivative of a function?'. The obvious starting point is to consider an analytic function $f(x)$, expressed as a power series $f(x)=\sum_{n} c_{n} x ^n$. The fractional derivative of $f(x)$ can then be obtained by deriving each term in the series, and hoping that the final series converge.
When $s$ is an integer, the $s$th derivative of the term $x^n$ is given by $(n!/(n-s)!) x^{n-s}$, which can be expressed in terms of gamma functions as $(\Gamma(n+1)/\Gamma(n-s+1)) x^{n-s}$ where now, however, $s$ is allowed to take real values. On the other hand, the $s$th-iterated integral of $f(x)$ can be computed from its Laplace transform: Let $I_{x}^1 = \int_{0}^x f(u) du$. Its Laplace transform is  
${\cal{L}}\{ I_{x}^1 \} = (1/\omega) {\cal{L}} \{f(x)\}$. Thus, the Laplace transform of the nth-iterated integral of $f(u)$ will be $(1/\omega^n){\cal{L}}\{f(x)\}$. Now, we take $n$ to be real $n\rightarrow \alpha$, i.e.,  ${\cal{L}}\{I_{x}^\alpha \}=(1/\omega^\alpha){\cal{L}}\{f(x)\}$, which is well-defined. By transforming back, we obtain $I_{x}^{\alpha}$ via the convolution 
$I_{x}^{\alpha} = (1/\Gamma(\alpha)) \int_{0}^{x} f(u)/(x-u)^{1-\alpha}$.

From these early attempts to define fractional derivatives and integrals,  the continuing efforts to create a consistent theory, have promoted fractional calculus from a mathematical curiosity to a full-fledged branch of mathematics. Nowadays, a well-used definition for the fractional derivative is the Caputo derivative:
\be
{d^s\over{d x^s}} f(x)={1\over{\Gamma(1-s)}} {d\over{d x}}\int_{0}^x {f'(u)\over{(x-u)^s}} d u,
\ee
where $0<s<1$.

The fractional form $(-\Delta)^s$ of the laplacian operator $\Delta=\partial^2/\partial {\bf r}^2$, can be expressed as\cite{landkof}
\be
(-\Delta)^s f({\bf x}) = 
{4^s \Gamma[(d/2)+s]\over{\pi^{d/2} |\Gamma(-s)|}}  \int { f({\bf x})-f({\bf y})\over{|{\bf x}-{\bf y}|^{d + 2 s}} }
\label{dos}
\ee
where $d$ is the dimension.

It has been found that fractional-order differential equations constitute useful tools to articulate complex events and to model various physical phenomena. In particular, the fractional Laplacian (\ref{dos}) has found many applications in fields as diverse as 
fluid mechanics\cite{26,27}, plasmas\cite{plasma}, fractional kinetics and anomalous diffusion\cite{71,86,101}, Levy processes in quantum mechanics\cite{75},  strange kinetics\cite{82},  biological invasions\cite{9}, fractional quantum mechanics\cite{64,65},  and electrical propagation in cardiac tissue\cite{20}. 

In this work, we are interested in the  interplay of disorder and fractionality in a simple discrete Anderson model, defined as a one-dimensional tight-binding model with random site energies extracted from a uniform distribution of width $W$ and endowed with a {\em discrete} version of the fractional Laplacian (\ref{dos}).  In particular, we are interested in studying how fractionality affects localization and the transport of excitations.
\vspace{0.1cm}

\noindent
{\bf 2.\ The model}.\\ 
Let us start with the well-known one-dimensional Anderson model ,
\be 
i {d C_{n}(z)\over{d t}} +\epsilon_{n} C_{n}(t) + V (C_{n+1}(t) + C_{n-1}(t))=0\label{eq:1}
\ee
where $\epsilon_{n}$ is extracted from a uniform random distribution $[-W, W]$. For a finite chain, Eq.(\ref{eq:1}) is valid for $1<n<N$, while at the edges we have
\be
i {d C_{1}(t)\over{d z}} + \epsilon_{1} C_{1}(t) + V\ C_{2}(t)=0
\ee
and 
\be
i {d C_{N}(t)\over{d z}} + \epsilon_{N} C_{N}(t) + V\ C_{N-1}(t)=0.
\ee

Equation (\ref{eq:1}) can be recast as 
\be 
i {d C_{n}(z)\over{d t}} -2 V +  \epsilon_{n} C_{n}(t) + (\Delta_{n}) C_{n}(t)=0\label{eq:6}
\ee
where $\Delta_{n}$ is the discrete Laplacian $\Delta_{n} C_{n} = C_{n+1} - 2 C_{n}+C_{n-1}$.
Now we proceed to replace this discrete Laplacian by the fractional discrete Laplacian\cite{roncal}
\be
(-\Delta_{n}^{s}) C_{n} =V \sum_{m\neq n} K^{s}(m-n) (C_{m}-C_{n}),\label{eq:2}
\ee
with the kernel 
\be
K^{s}(n) = {4^s \Gamma(s+(1/2))\over{\sqrt{\pi} |\Gamma(-s)|}} {\Gamma(|n|-s)\over{\Gamma(|n|+1+s)}}, \label{eq:3}
\ee
where $\Gamma(x)$ is the gamma function. Figure 1 shows the behavior of $K^s(m)$ as a function of $s$ for several coupling distance values $m$. In the limit $s\rightarrow 1$, only the nearest-neighbor coupling is different from zero: $K^s(m)\approx \delta_{m\pm 1}$, while for $s\rightarrow 0$, $K^s(m)\approx 0$ for all $m$. At long distances, $K^s(m)\rightarrow 1/|m|^{1+2 s}$. Except for $m=1$, the value of all couplings $K^s(m)$ are zero at both extremes $s=0$ and $s=1$, with a non-symmetrical maximum value in between.

After we replace Eqs.(\ref{eq:2}), (\ref{eq:3}) in  (\ref{eq:1}) and after posing a stationary-state solution $C_{n}(t) = \exp(i\lambda t) \phi_{n}$ we  obtain a set of nonlinear difference equations for $\{\phi_{n}\}$:
\be
(-\lambda + 2 V+\epsilon_{n})\phi_{n} + V \sum_{m\neq n} K^{s}(m-n) (\phi_{m}-\phi_{n})=0\label{eq:4}
\ee
where, for $n=1$ and $n=N$ we must replace the $2 V$ term in Eq.(\ref{eq:4}) by $V$. 

In the absence of disorder ($\epsilon_{n}=0$) and after we pose a plane-wave solution $\phi_{n} \sim \exp(i k n)$ we obtain, 
\be
\lambda(k) = 2 V - 4 V \sum_{m=1}^{\infty} K^{s}(m) \sin((1/2) m k)^2.
\ee
In a previous work\cite{previous} we have found an exact form for $\lambda(k)$ from which we were able to obtain the behavior of the bandwidth and the mean square displacement, as a function of the fractional exponent $s$, in closed form. In our case here, and due to the presence of disorder ($\epsilon_{n}\neq 0$), we will pursue a more numerical approach. 
\begin{figure}[t]
 \includegraphics[scale=0.25]{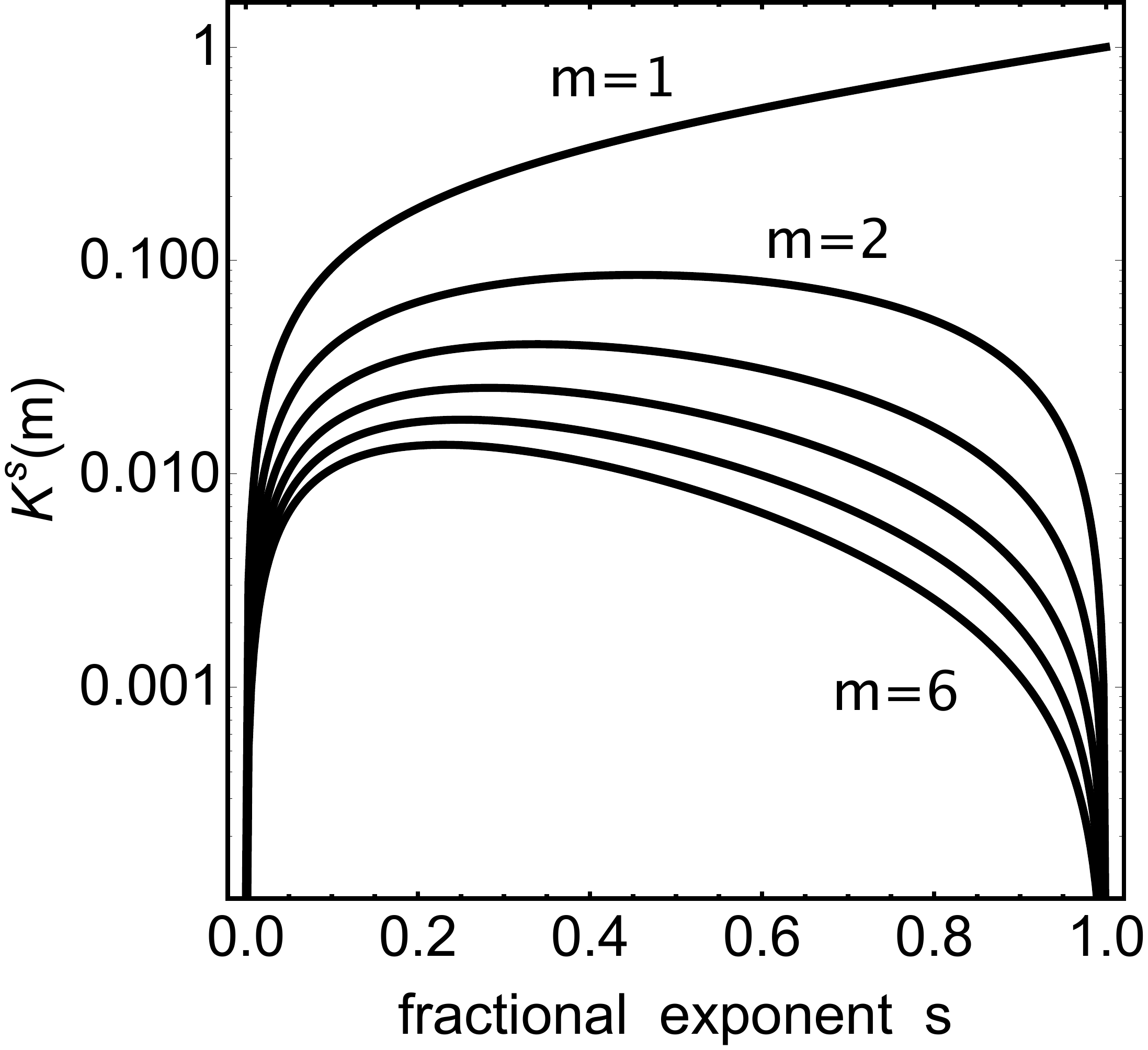}
  \caption{Coupling $K^{s}(m)$ as a function of the fractional exponent $s$ for several coupling distances.}
  \label{fig0}
\end{figure}
\vspace{0.5cm}

\noindent
{\bf 3.\ Results}.\\
We start by producing a random sequence of site energies $\epsilon_{n}$ extracted from a uniform random distribution, $\epsilon_{n}\in[-W,W]$. By straightforward diagonalization 
we compute the eigenvalues $\{\lambda\}$ corresponding with a given fractional exponent $s$, going from $s=1$ down to $s=0$. Results are shown in Fig. 2 where, for a given $s$ value and a given $\epsilon_{n}$ sequence, we mark all allowed eigenvalues with a black dot. 
\begin{figure}[t]
 \includegraphics[scale=0.22]{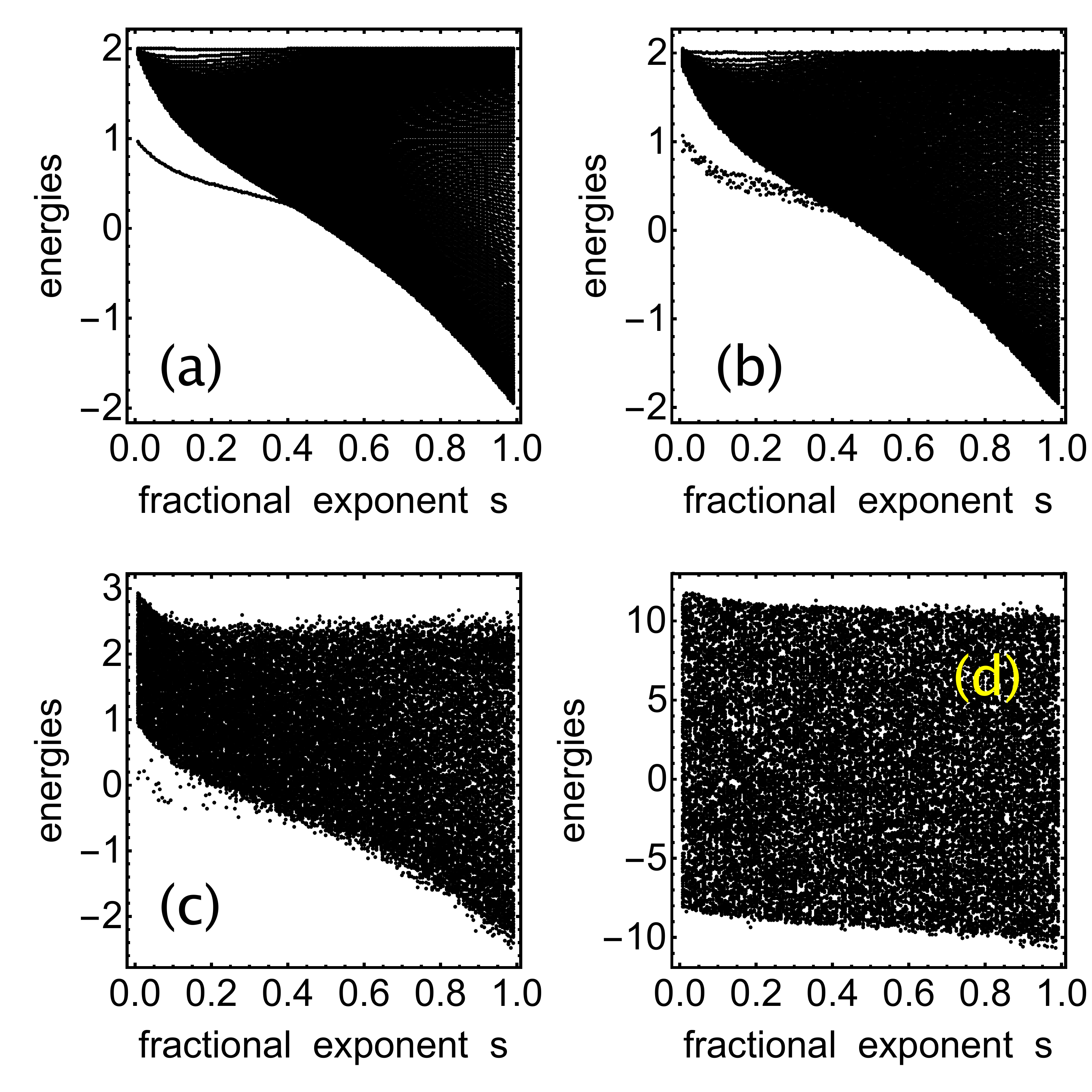}
  \caption{Energies $\lambda$ of the disordered chain as a function of the fractional   exponent $s$ for several disorder widths: (a) $W=0$, (b) $W=0.1$, (c) $W=1$ (d) $W=10$. ($N=133$)}
  \label{fig1}
\end{figure}
In the absence of disorder ($W=0$), the main feature we notice is that the bandwidth shrinks with decreasing $s$ (plot 1(a)). At $s=0$ the bandwidth is precisely zero, marking the onset of a flat band. Another interesting behavior is that at some $s$ value (approximately  $0.415$), the two lowest eigenvalues, detach themselves from the main band, reaching a common value of $V$ as $s\rightarrow 0$. Thus, a gap is formed between the band and the ground state. This is reminiscent of the behavior observed in a discrete Anderson model with an infinite-range hopping, in the no-disorder limit\cite{celardo}. In the $s\rightarrow 0$ limit, the form of the two degenerate eigenstates can be easily obtained from Eq.(\ref{eq:4}):
\begin{eqnarray}
(-\lambda + V) \phi_{1}&\approx &0\hspace{1cm} n=1\nonumber\\
(-\lambda + V) \phi_{N}&\approx &0\hspace{1cm} n=N\nonumber\\
(-\lambda + 2V) \phi_{n}&\approx &0\hspace{1cm} n\neq 1,N\label{eq:5}
\end{eqnarray}
Now, since the ground state energy is $\lambda=V$ for $s=0$, we deduce from Eq.(\ref{eq:5})
that $\phi_{1} =\phi_{N}\approx (1/\sqrt{2})$ and $\phi_{1}=-\phi_{N}\approx(1/\sqrt{2})$, while $\phi_{n}\approx 0$ for $n\neq 1,N$. For the flat band case, $\lambda=2 V$, we have
$\phi_{1}=0=\phi_{N}$ and for the remaining $N-2$ states, $\phi_{n}\neq 0$ for $1<n<N$. In fact, since in this limit all sites are decoupled from each other, the orthonormalized eigenvectors have the form $(0,1,0,\cdots,0), (0,0,1,0,\cdots,0), \cdots, (0,0,\cdots,1,0)$.

In the presence of disorder ($W\neq 0$), the degenerate `tongue' becomes more and more blurred as the disorder width is increased, leading to an `evaporation' of the `tongue' and later its complete merging with the rest of the eigenvalues. If we look at the $s\rightarrow 0$ limit, we note that, because the coupling $K^s(m)\rightarrow 0$, the stationary equations now read $(-\lambda+2V(V)+\epsilon_{n}) \phi_{n}\approx 0$, where $V$ is to be used for $n=1, N$ and $2 V$ for the rest. This implies $\lambda^{(n)}=\{2 V (V) + \epsilon_{n}\}$ with eigenvectors $\{\phi_{m}^{\lambda^{(n)}}\}=\delta_{m, n}$, a set of fully-decoupled localized modes. 

In Fig.3 we show the realization-averaged density of states
\be
\Omega(E)=\left\langle {1\over{N}} \sum_{\alpha} \delta(E - E_{\alpha})\right\rangle
\ee
where $E_{\alpha}$ is the energy of the $\alpha$th state. We plot $\Omega(E)$ for several different disorder widths and fractional exponents.  For a fixed disorder width, we see that as $s$ is decreased the width of the density of states decreases as well. In fact, for the case $W=0$ and small $s$, $\Omega(E)$ shows a discrete peak near the lower edge of the band. This is precisely the energy of the two degenerate states we mentioned before. As disorder increases, this extra peak is quickly lost. This shrinking of the point spectrum with decreasing $s$ is observed for all disorder widths examined. On the other hand, for a fixed fractional exponent, an increase in disorder width $W$ causes $\Omega(E)$ to widen. For the standard, non-fractional case ($s=1$) this is well-known, but it seems to hold for any value of the fractional exponent as well. Therefore, one might counterbalance the widening of $\Omega(E)$ caused by an increase in $W$, with a judicious decreasing in $\Omega(E) $caused by a decrease in $s$.
\begin{figure}[t]
 \includegraphics[scale=0.18]{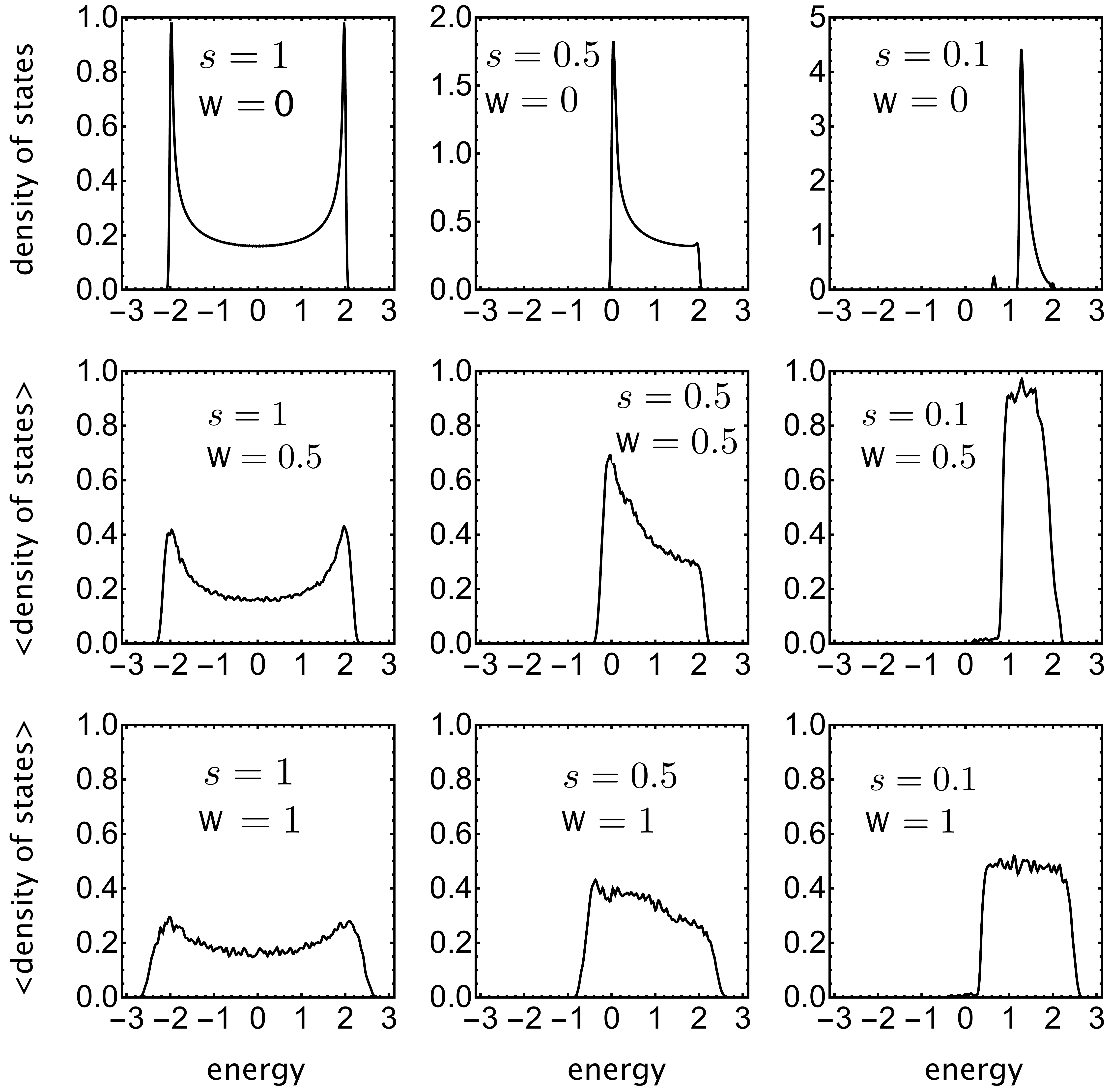}
  \caption{Realization-averaged density of states $\Omega(E)$ for several fractional exponent values $s$  and disorder widths $W$ ($N=133, \mbox{number of realizations}=133$).}
  \label{fig2}
\end{figure}
\begin{figure}[b]
 \includegraphics[scale=0.3]{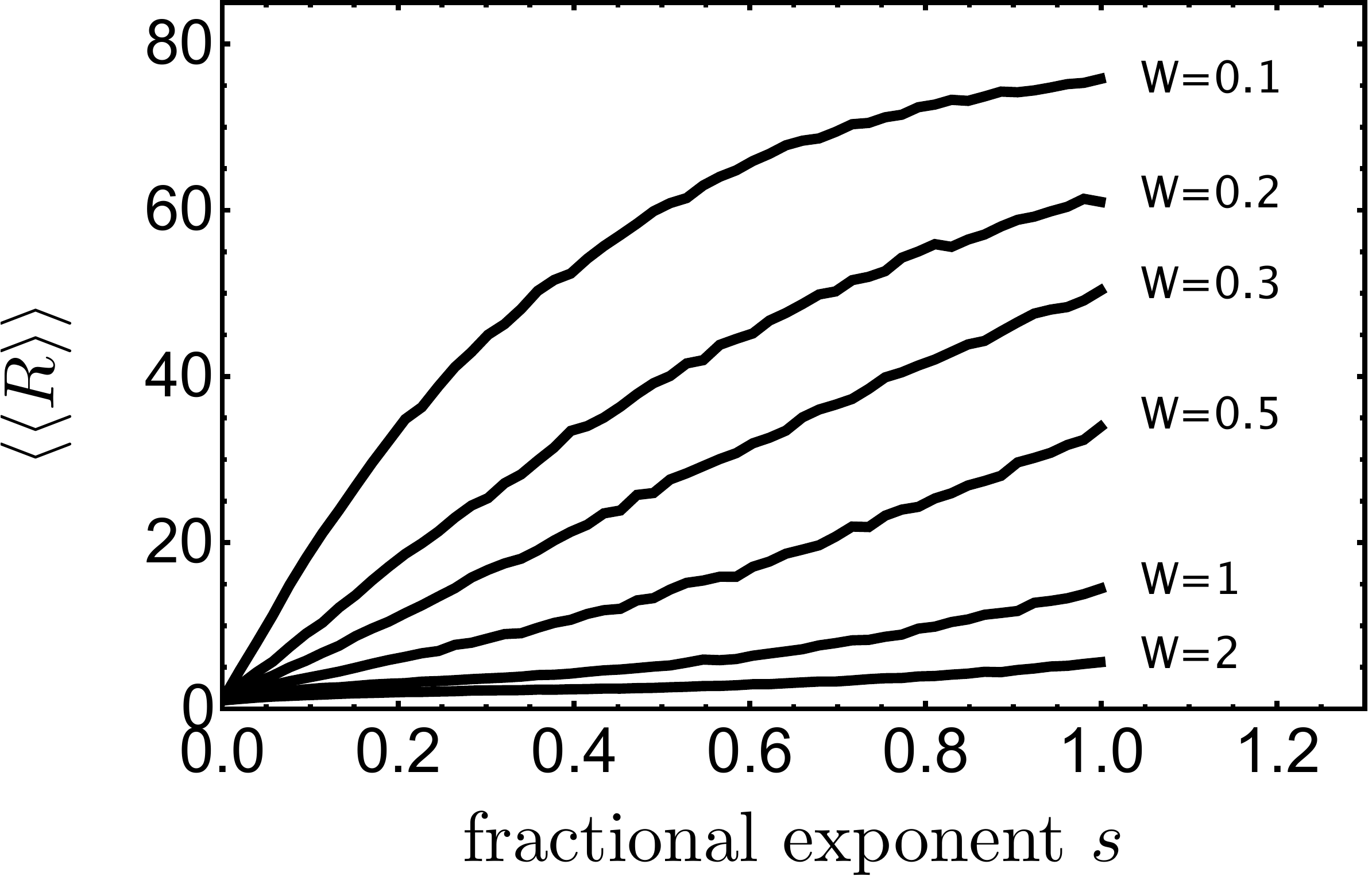}
  \caption{Realization-and state average participation ratio as a function of the fractional exponent, for several disorder widths ($N=133$, $\mbox{number of realizations}=33$)}
  \label{fig1}
\end{figure}
A useful indicator of localization of a state $\{\phi_{n}\}$ is the participation ratio $R$, defined as
\be
R = {(\sum_{n} |\phi_{n}|^2 )^2\over{\sum_{n} |\phi_{n}|^4}}.
\ee
For a delocalized state, $R\rightarrow N$ while for a completely localized one, $R\rightarrow 1$. We compute the disorder-and mode average of $R$ and plot it as a function of the fractional exponent $s$. Results are shown in Fig.4 that shows $\langle\langle R\rangle\rangle$ as a function of $s$ for several disorder widths. As can be clearly seen, for a given disorder strength $W$, a decrease in $s$ reduces $\langle\langle R\rangle\rangle$, meaning a higher degree of localization. For a fixed fractional exponent $s$, an increase in $W$ also decreases $\langle\langle R\rangle\rangle$ signaling a greater localization. To summarize: A deviation of $s$ from its standard value ($s=1$) increases the average localization.

While the average participation ratio $\langle\langle R\rangle\rangle$ is useful to have an idea of the tendency towards localization, a dynamical measure of the propagation of an excitation is given by the mean square displacement (MSD), defined as
\be
\mbox {MSD} = {\sum_{n}n^2 |C_{n}(t)|^2\over{\sum_{n} |C_{n}(t)|^2}}
\ee
where $C_{n}(0)=\delta_{n,0}$ and $C_{n}(t)$ is the solution to Eq.(\ref{eq:1}).

Figures 5(a),(b) show results for the time evolution of the realization-averaged 
$\langle MSD\rangle$, for several $s$ values and fixed disorder width $W=1$. We identify three $s$ regimes: In the first one, from $s=1$ down to $s\sim 0.4$, the $\langle MSD\rangle$ shows a tendency towards saturation as a function of time. It also decreases with decreasing $s$ meaning that propagation is more inhibited. In the second regime, 
$0.15 < s <0.4$, this tendency is reversed and now there is no hint of saturation in time while $\langle MSD\rangle$ increases with decreasing $s$. We can appreciate a distinct change of slope in $\langle MSD\rangle$. Thus, in this regime propagation is enhanced with decreasing $s$. Finally, for $0 < s < 0.15$, there is a precipitous fall of $\langle MSD\rangle$ towards zero as $s$ approaches zero.

To have a more clear picture, we plot in Figs.5(c),(d),(e),(f) the average $\langle MSD\rangle$ at some fixed time $t=t_{max}$, normalized to $t_{max}^2$, for convenience. This corresponds to plotting the endpoints of the curves shown in Figs.5(a),(b). In the first case (Fig.5c), we take $W=0$ and plot $MSD(t_{max})/t_{max}^2$ as a function of the fractional exponent. Since in this case the propagation is always ballistic\cite{previous}, the quantity $\langle MSD\rangle/t_{max}^2$   corresponds to the dimensionless ballistics `speed'. We see a monotonic decrease of this speed towards zero, which is tempered a bit around $s=0.3$ where there is an inflection point. This can be explained by looking at the shape of the coupling $K(m,s)$ as a function of $s$  for various coupling distances $m$ (Fig.1). In the vicinity of $s=1$, the coupling is dominated by nearest-neighbor interactions K(1,s). As this coupling decreases so does the MSD.  As $s$ is decreased further, the rest of the couplings $\{K(2,s), K(3,s),\cdots,\}$ `kick in', and reduce the falling rate of the MSD. Finally, in the vicinity of $s=0$, all the $\{K(m,s)\}$ approach zero and the falling rate of the MSD resumes, reaching zero at $s=0$.

When we add disorder, the picture changes and now, after an initial decrease, it develops a a `hump' in the low $s$ sector which falls precipitously to zero as $s\rightarrow 0$. This behavior is observed at small ($W=0.5$), medium ($W=1$) and large ($W=2$) disorder widths (Figs. d,e,f).\\
 
\begin{figure}[t]
 \includegraphics[scale=0.25]{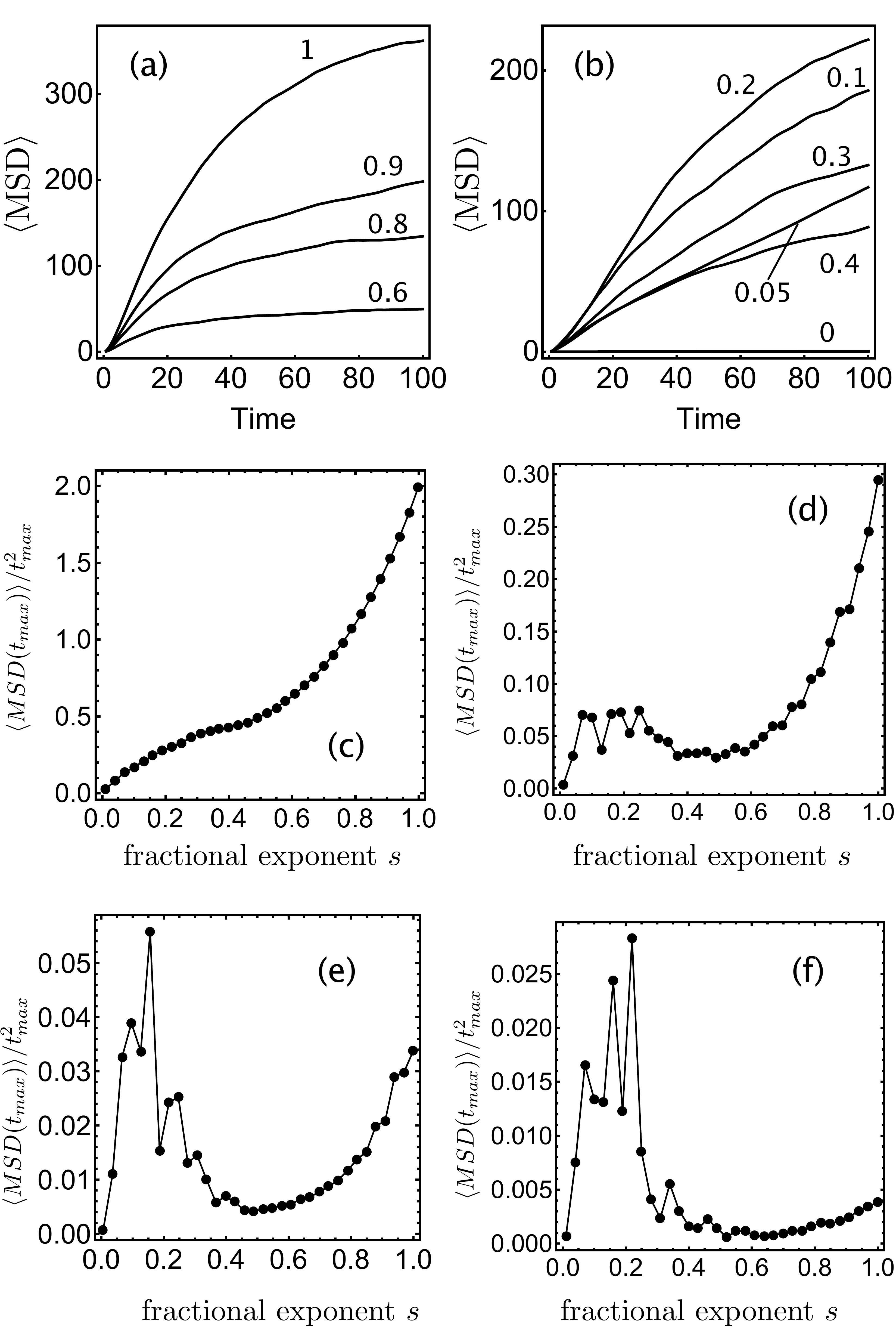}
  \caption{Upper row: Average mean square displacement as a function of time, for disorder width $W=1$, computed for $N=600$, $100$ random realizations, and for times up to $t=100$. The numbers on each curve denote the value of the fractional exponent. Middle and lower rows: Average mean square displacement at a fixed time $t_{max}$ as a function of the fractional exponent, for a fixed disorder width $W=0$ (c), $W=0.5$ (d), $W=1$ (e) and $W=2$ (f) ($N=600, t_{max}=100$, number of realizations $=33$).}
  \label{fig4}
\end{figure}
\vspace{0.5cm}

\noindent
{\bf 4.\ Conclusions}.\\
We have examined the localization properties of a simple Anderson model in the simultaneous presence of diagonal disorder and a discrete fractional Laplacian. In the absence of disorder, we observe a narrowing of the point spectrum with decreasing fractional exponent. As $s$ decreases, the two lowest modes detach themselves from the band and, at $s\rightarrow 0$, they converge to $V$. At the same time, the remaining $(N-2)$ in the band converges to a  single energy $2 V$. In this way, a gap is formed between the ground state and a degenerate flat band. The ground state amplitude was computed analytically, as well as the shape of the flat band modes. For small and large disorder width, the point spectrum also narrows with decreasing $s$ but degeneracy is lifted, and a blurred tongue is still appreciable at small disorder, while at large disorder the energy tongue `evaporates' and merges with the wide spectrum. The density of states (DOS) widens with increasing disorder, while it shrinks with decreasing exponent. Thus, it is possible in principle, to counteract the widening of the spectrum with a judicious choice of the fractional exponent. 

The mode-and-disorder average of the participation ratio $\langle\langle R\rangle\rangle$ shows a monotonic decrease with an increase in width disorder and a decrease of the fractional exponent, pointing to a decrease of the average localization length. This is not the whole story, however. See below.

The dynamics of an initially localized excitation, measured with the mean square displacement (MSD), shows an interesting behavior. For the ordered case $W=0$, the MSD decreases monotonically with decreasing $s$, reaching a zero value at $s=0$, which is understandable since, at that value, we have a flat band, which in turn implies null mobility. As disorder width is increased, the mobility decreases considerably because of disorder but we also observe the onset of an interval in fractional exponent $s$ where the MSD increases, reaches a maximum value, to finally drop precipitously to zero as $s$ approaches zero. That is, there is an interval where the already small mobility increases with decreasing fractionality. This is puzzling since localization always increases with decreasing $s$. A possible explanation could lie in the observation that $K(m)\rightarrow 1/|m|^{1+2s}$ at large $m$, which implies that we have an effective long-range hopping at small $s$. This is reminiscent of the Anderson model with a long-range hopping, where it was proved that a population of extended states can exist\cite{adame,brito}. In our case, it is this fraction of the total population that is responsible for the growth of the MSD as $s$ decreases around $s\sim 0.4 - 0.2$. It seems to be a small population fraction since it does not show up in $\langle\langle R\rangle\rangle$. Later, as we bring $s$ closer to zero, the overall coupling decreases quickly, leading to a vanishing MSD at $s=0$. 

Therefore, the main effect of fractality $s$ is a shrinking of the point spectrum and of the density of states, and an enhancement of the average localization length, save for a small population of extended states at small $s$ values, that is responsible for a degree of propagation. Our fractional Anderson model seems akin to  a long-range hopping Anderson model.
\vspace{0.5cm}

{\bf Acknowledgments}

This work was supported by Fondecyt Grant 1200120.

\end{document}